\documentclass[aps,prc,letterpaper,11pt,twoside,tightenlines,nofootinbib,showpacs,preprint]{revtex4-1}
\usepackage{graphicx}
\usepackage[sort&compress]{natbib}
\usepackage{subfigure}
\usepackage{amsmath}
\usepackage{amsfonts}
\usepackage{cancel}

\begin{document}

\arraycolsep1.5pt

\newcommand{\Ima}{\textrm{Im}}
\newcommand{\Rea}{\textrm{Re}}
\newcommand{\mev}{\textrm{ MeV}}
\newcommand{\be}{\begin{equation}}
\newcommand{\ee}{\end{equation}}
\newcommand{\ba}{\begin{eqnarray}}
\newcommand{\ea}{\end{eqnarray}}
\newcommand{\gev}{\textrm{ GeV}}
\newcommand{\nn}{{\nonumber}}
\newcommand{\dtres}{d^{\hspace{0.1mm} 3}\hspace{-0.5mm}}

\title{Transparency ratio in $\gamma A \to \eta^\prime A^\prime$ and the in-medium $\eta^\prime$ width}

\author{M. Nanova$^{1}$, V. Metag$^{1}$, A. Ramos$^{2}$, E. Oset$^{3}$, I. Jaegle$^{4,a}$, K. Makonyi$^{1}$,  K. Brinkmann$^{5}$, O. Bartholomy$^5$, D.~Bayadilov$^{5,6}$, Y.A.~Beloglazov$^6$, V.~Crede$^{5,b}$, H.~Dutz$^5$, A.~Ehmanns$^5$, D.~Elsner$^7$, K.~Essig$^5$, R.~Ewald$^7$,~I.~Fabry$^5$, M.~Fuchs$^5$, Ch.~Funke$^5$, R.~Gregor$^1$, A.B.~Gridnev$^6$, E.~Gutz$^5$, S.~H\"offgen$^5$,~P.~Hoffmeister$^5$, I.~Horn$^5$, J.~Junkersfeld$^5$, H.~Kalinowsky$^5$,~Frank~Klein$^7$,~Friedrich~Klein$^7$, E.~Klempt$^5$, M.~Konrad$^7$, B.~Kopf$^{8,9}$, B.~Krusche$^4$, J.~Langheinrich$^{7,9}$, H.~L\"ohner$^{10}$, I.V.~Lopatin$^6$, J.~Lotz$^5$, S.~Lugert$^1$, D.~Menze$^7$, T.~Mertens$^4$, J.G. Messchendorp$^{10}$, C.~Morales$^7$,~R.~Novotny$^1$, M.~Ostrick$^{7,c}$, L.M.~Pant$^{1,d}$, H.~van Pee$^5$, M.~Pfeiffer$^1$, A. Roy$^{1,e}$, A.~Radkov$^6$,~S.~Schadmand$^{1,f}$, Ch.~Schmidt$^5$, H.~Schmieden$^7$, B.~Schoch$^7$,~A.~S\"ule$^7$, V.~V.~Sumachev$^6$, T.~Szczepanek$^5$, U.~Thoma$^5$, D.~Trnka$^1$, R.~Varma$^{1,g}$, D.~Walther$^5$,  Ch.~Wendel$^5$\\
(The CBELSA/TAPS Collaboration)}
\affiliation {
{$^{1}$ II. Physikalisches Institut, Universit\"at Gie{\ss}en, Germany}\\
{$^{2}$ Departament d'Estructura i Constituents de la Materia, Universitat de Barcelona, Spain}\\
{$^{3}$ Departamento de F\'{\i}sica Te\'orica and IFIC, Centro Mixto Universidad de Valencia-CSIC, Institutos de Investigaci\'on de Paterna, Aptdo. 22085, 46071 Valencia, Spain}\\
{$^{4}$Physikalisches Institut, Universit\"at Basel, Switzerland}\\
{$^{5}$Helmholtz-Institut f\"ur Strahlen- u. Kernphysik Universit\"at Bonn, Germany }\\
{$^6$Petersburg Nuclear Physics Institute Gatchina, Russia}\\
{$^{7}$Physikalisches Institut, Universit\"at Bonn, Germany}\\
{$^{8}$Institut f\"ur Kern- und Teilchenphysik TU Dresden, Germany}\\
{$^{9}$Physikalisches Institut, Universit\"at Bochum, Germany}\\
{$^{10}$Kernfysisch Versneller Institut  Groningen, The Netherlands}\\
{$^{a}$Current address: Hawaii University, USA}\\
{$^{b}$Current address: Florida State University Tallahassee, FL, USA}\\
{$^{c}$Current address: Physikalisches Institut Universit\"at Mainz, Germany}\\
{$^{d}$Current address: Nuclear Physics Division BARC, Mumbai, India}\\
{$^{e}$Current address: IIT Indore, Indore, India}\\
{$^{f}$Current address: Forschungszentrum J\"ulich, Germany}\\
{$^{g}$Current address: Department of Physics I.I.T. Powai, Mumbai, India}\\
}
\date{\today}
\begin{abstract} 
 The photoproduction of $\eta^\prime$-mesons off different nuclei has been measured with the CBELSA/TAPS detector system for incident photon energies between 1500 - 2200 MeV. The transparency ratio has been deduced and compared to theoretical calculations describing the propagation of $\eta^\prime $-mesons in nuclei.  The comparison indicates a width of the $\eta^\prime $-meson  of the order of $\Gamma= 15-25$ MeV at $\rho =\rho_0$ for an average momentum $p_{\eta^\prime} $ = 1050 MeV/c, at which the $\eta^\prime$-meson is produced in the nuclear rest frame. The inelastic  $\eta^\prime N$ cross section is estimated to be 3 - 10~mb. Parameterizing the photoproduction cross section of $\eta^\prime $-mesons by $\sigma(A) = \sigma_0 A^{\alpha}$, a value of $\alpha  = 0.84 \pm 0.03$ has been deduced.
 \end{abstract}
\maketitle

\section{Introduction}
\label{Intro}
The $\eta^\prime$-meson has interesting 
properties concerning the underlying QCD dynamics 
of hadrons which are related to the $U_A (1) $ axial vector anomaly \cite{kogut,weinberg,tooft,witten}. Being close to a singlet of SU(3), its interaction with nucleons is supposed to be weak compared for instance with the case of its partner the $\eta$. 

The  $\eta^\prime N$ scattering length has been estimated from the study of the $pp \to pp \eta^\prime$ cross section near threshold at COSY  \cite{Moskal:2000gj,Moskal:2000pu}. A refined analysis of this reaction, comparing the cross section with that of the $pp \to pp \pi ^0$ reaction, concluded that the scattering length should be of the order of magnitude of that of the $\pi N$ interaction and hence $|a_{\eta^\prime N}|\sim 0.1$ fm \cite{Moskal:2000pu}. This indicates a rather weak $\eta^\prime N$ interaction. In \cite{Oset:2010ub} it was interpreted as a consequence of the particular dynamics of the singlet of mesons together with a small admixture of the $\eta^\prime$ with the octet. This is related to the mixing angle of u,d  and strange quarks in the $\eta$ and $\eta^\prime$~\cite{Gilman:1987ax,Bramon:1989kk,Schechter:1992iz,Leutwyler:1996sa,Bramon:1997va,Mathieu:2010ss}, indicating that the $\eta^\prime N$ amplitude is sensitive to this mixing angle.\\

Another experimental approach to learn more about the $\eta^\prime N$ interaction is the study of $\eta^\prime $ photoproduction off nuclei which provides information on in-medium properties of the $\eta^\prime $-meson. The in-medium width of the $\eta^\prime $-meson can be extracted from the attenuation of the $\eta^\prime$-meson flux deduced from a measurement of the transparency ratio for a number of nuclei. Unless when removed by inelastic channels the $\eta^\prime$-meson will decay outside of the nucleus because of its long lifetime and thus its in-medium mass is not accessible experimentally. Recently, however, indirect evidence has been claimed for a dropping 
$\eta^\prime$ mass in the hot and dense matter formed in ultrarelativistic heavy-ion collisions at RHIC energies \cite{Csoergo}. The in-medium width provides information on the strength of the $\eta^\prime N$ interaction and it will be instructive to compare this result with in-medium widths obtained for other mesons (see section C). Furthermore, knowledge of the $\eta^\prime$ in-medium width is important for the feasibility of observing $\eta^\prime$-nucleus bound systems, theoretically predicted in some models~\cite{Nagahiro,Nagahiro1}.

\section{Transparency ratio in $\eta^\prime$ photoproduction}

\subsection{Formalism}
 It was shown in \cite{eli} that by comparing photoproduction cross sections on different nuclei one could extract
 the widths in nuclear matter of the produced  particles.  The relevant magnitude in this comparison is the transparency ratio, or ratio of production cross sections per nucleon in different nuclei with respect to the elementary cross section on the nucleon. The photoproduction cross section $A (\gamma,\eta^\prime)A^\prime$ in nuclei is not proportional to $A$ for different nuclei, and the 
 deviation  from the $A$ scaling can be related to the width of the produced particle in the nucleus. The formalism is straightforward  \cite{eli,phimedpro,murat} and one does not need a particular model for the elementary production process. For the case of photoproduction, we only need to use the fact that the photon probability to produce a primary  $\eta^\prime$ in a volume $d^3 r$ is proportional to the number of nucleons in this volume, $\rho(r) d^3r$. The number of $\eta^\prime$-mesons that survives without absorption and leaves the nucleus is proportional to the survival probability
\begin{eqnarray}
&&P_s(\vec{k}_{\eta^\prime},\vec{r}\,) = {\rm exp}\left[{\int_0^\infty dl \frac{{\rm Im}\, \Pi_{\eta^\prime}(\rho(\vec{r}\,^\prime))}{\mid \vec{k}_{\eta^\prime}\mid}}\right]\ , \\
&&~~~~{\rm with}~~~\vec{r}\,^\prime= \vec{r}+ l \frac{\vec{k}_{\eta^\prime}}{\mid \vec{k}_{\eta^\prime}\mid} \ , \nonumber
\label{eq:prob}
\end{eqnarray}
where $\Pi_{\eta^\prime}$ is the selfenergy of the $\eta^\prime$ in the nucleus, $\vec{r}$ the production point and 
$\vec{k}_{\eta^\prime}$ the $\eta^\prime$ momentum in the lab frame, the direction of which is determined according to the experimental differential cross sections at the corresponding incoming photon energy~\cite{Crede:2009zzb}.
  Hence, the cross section for quasifree photoproduction of $\eta^\prime$-mesons in the nucleus is given by 
\begin{equation}
\sigma_{\gamma A\to \eta^\prime A^\prime}= C \int d^3r \rho(\vec{r}\,)
\int_0^{2\pi} d(\phi^{\eta^\prime}_{\rm c.m.}) \int_{-1}^{1} d(\cos{\theta^{\eta^\prime}_{\rm c.m.}}) \frac{d\sigma}{d\Omega}(\gamma p \to \eta^\prime p) 
P_s(\vec{k}_{\eta^\prime},\vec{r}\,)  \ ,
\label{eq:sigma}
\end{equation}
 and the transparency ratio for a given nucleus is given by
\begin{equation}
\tilde{T}_A=\frac{\sigma_{\gamma A\to \eta^\prime A^\prime}}{A
\sigma_{\gamma N\to \eta^\prime N} } \ .
\label{eq:trans}
\end{equation} 
Here, the production cross section per nucleon within a nucleus is compared to the production cross section on a free nucleon which is a measure for the absorption of the $\eta^\prime$ within the nucleus. $A$ is hereby the effective number of participant nucleons reached by the photon beam which decreases due to photon shadowing relative to the total number of available nucleons with incident photon energy and target size. As shown in \cite{bianchi}, at the average photon energies of our experiment the shadowing of the photons results in an effective number of participant nucleons per nucleon of 0.88 for C and 0.84 for Pb. There is a difference of 5 $\%$ from C to Pb. This means that in the transparency ratio of eq. (4), 5 $\%$ of the decrease of this ratio from C to Pb is due to the shadowing of the photons in the initial photon propagation and not to the absorption of the $\eta^\prime$ in the final state interaction with the nucleus. To correct for this we increase the measured ratio of eq. (4) by 5$\%$ for Pb and correspondingly  by 2 $\%$ for Ni and 1$\%$ for Ca, taking C as reference.

Furthermore, Eqs. (1) and (2) rely upon a single step process for $\eta^\prime$ production, i.e. the elementary reaction $\gamma p \rightarrow \eta^\prime p$.  We shall provide experimental support later on for the smallness of the two step mechanisms, but a justification can also be found theoretically by the fact that the usual steps: $\gamma N \rightarrow \pi N$, followed by $\pi N -> \eta^\prime N$ are practically negligible, given the abnormally small  $\pi N \rightarrow \eta^\prime N$ cross section of the order of 0.1mb \cite{Landolt-Boernstein}. Furthermore, as shown in  Ref.~\cite{phimedpro}, there is an additional reduction of the effects of the multistep processes if one considers the transparency ratio relative to that of a medium-light nucleus. We thus take $^{12}$C as the nucleus of reference and will evaluate the ratio
 \begin{equation}
T_A=\frac{\tilde{T}_A}{\tilde{T}_{12}} \ .
\label{eq:trans_ratio}
\end{equation} 

It is clear that a measurement of the transparency ratio, in the form of Eq.(\ref{eq:trans}) or expressed as a fraction of the $^{12}$C transparency ratio as in Eq.(\ref{eq:trans_ratio}), provides information on the $\eta^\prime$ self-energy in a medium or, alternatively, its width:
\begin{equation}
\Gamma_{\eta^\prime}(\rho)=-\frac{{\rm Im}\, \Pi_{\eta^\prime}(\rho)}{E_{\eta^\prime}} \ ,
\label{eq:gamma}
\end{equation}
where $E_{\eta^\prime}$ is the $\eta^\prime$ energy in the lab frame.
The low-density theorem, which can be applied because the $\eta^\prime N$ amplitude is rather small,
\begin{equation}
\Pi_{\eta^\prime}(\rho)= t_{\eta^\prime N \to \eta^\prime N}   \rho \ ,
\label{eq:trho}
\end{equation}
allows us to write
\begin{equation}
\Gamma_{\eta^\prime}(\rho)=\Gamma_{\eta^\prime}(\rho_0)\frac{\rho}{\rho_0}
 \ ,
\label{eq:gamma_rho}
\end{equation}
where $\rho_0$ can be taken as normal nuclear matter density, $\rho_0=0.17$ fm$^{-3}$. Using the local density approximation, which was shown to be exact for an s-wave amplitude 
in \cite{Nieves:1993ev}, one may
substitute the density of an infinite medium, $\rho$, by the actual density profile, $\rho(r)$,  of the nucleus, which we take from 
experiment  \cite{tables}.  In this way we obtain, via Eqs. (\ref{eq:prob}) to (\ref{eq:trans_ratio}), the transparency ratios for a set of nuclei ($^{12}$C, $^{16}$O, $^{24}$Mg, $^{27}$Al, $^{28}$Si, $^{31}$P, $^{32}$S, $^{40}$Ca, $^{56}$Fe, $^{64}$Cu, $^{89}$Y, $^{110}$Cd, $^{152}$Sm, $^{208}$Pb and $^{238}$U)
starting from different values of $\Gamma_{\eta^\prime}(\rho_0)$. By comparing with experiment, one may then obtain information on the $\eta^\prime N$ scattering amplitude as seen from Eq.(\ref{eq:trho}).
However, there is a caveat that we must take into account when analyzing the survival probability to obtain the quantity ${\rm Im}\,t_{\eta^\prime N \to \eta^\prime N}$. Indeed, ${\rm Im}\,t_{\eta^\prime N \to \eta^\prime N} $ is related to the $\eta^\prime N$ cross section via the optical theorem, which in our normalization stands as 
  \begin{equation}	 
{\rm Im}\,t_{\eta^\prime N \to \eta^\prime N} = -\frac{2 p^{\eta^\prime}_{\rm c.m.} \sqrt{s}}{2 M_N} \sigma_{\rm tot} \ ,
 \end{equation} 
where $p^{\eta^\prime}_{\rm c.m.}$ is the $\eta^\prime$ momentum and $s$ the square of the total energy in the $\eta^\prime N$ c.m. frame. 
Note that $\sigma _{\rm tot}$ contains the contribution of the reaction channels, where the $\eta^\prime$ disappears, as well as the integrated elastic cross section of $\eta^\prime N \to \eta^\prime N$. In the process, once the $\eta^\prime$ is produced in the first step, it will later collide with other nucleons. If the $\eta^\prime$ undergoes an inelastic process the $\eta^\prime$ will disappear from the flux and will be eliminated by means of the survival probability factor. Yet, if the $\eta^\prime$ undergoes an elastic collision (quasielastic in the nucleus), then it will change momentum and direction but will still be there and can be detected experimentally. This means that in measurements of the transparency ratio one obtains information on the reaction cross section, not $\sigma_{\rm tot}$. In other words, one would not be determining the complete imaginary part of the $t_{\eta^\prime N \to \eta^\prime N}$ amplitude but only the contribution coming from the inelastic channels.
 Note also that we are ignoring here the possibility of two or more nucleon induced absorption mechanisms. Thus, one should keep in mind that, while the in-medium width determined from transparency ratio data is a real measure of the absorption probability of the $\eta^\prime$ in the nucleus, its relationship to $\sigma_{\rm inel}$, the cross section for one-nucleon induced inelastic processes, is not straightforward. It is also usual to talk of an in-medium cross section  \cite{Ishikawa:2004id,Hartmann:2010zzb}, but this concept is not well suited for the case when part of the width comes from two or more nucleon induced  $\eta^\prime$ absorption. Recent calculations based upon the work of \cite{Oset:2010ub} indicate that the $\eta^\prime$ two nucleon induced absorption is relatively small \cite {Nagahiro1}. This allows us to determine an approximate $\eta^\prime N$ inelastic cross section by means of Eq.(\ref{eq:trho}). 
 However, there are also large uncertainties in the results of \cite {Nagahiro1} and we shall take them into account to quantify the uncertainties in the determination of  $\sigma_{\rm inel}$.

\subsection{Experiment and data analysis}

The experiment was performed at the electron stretcher accelerator in Bonn, using the combined Crystal Barrel(CB) and TAPS detectors which covered 99$\%$ of the full solid angle. Tagged photons with energies of 0.9 - 2.2 GeV, produced via bremsstrahlung at a rate of 8-10 MHz, impinged on a solid target. For the measurements, ${}^{12}\textrm{C}, {}^{40}\textrm{Ca}, {}^{93}\textrm{Nb}$ and ${}^{208}\textrm{Pb}$ targets were used with thicknesses of 20, 10, 1, and 0.6 mm, respectively, each corresponding roughly to about 8-10$\%$ of a radiation length. The data were collected during two running periods totaling 575~h. Events with $\eta^\prime $ candidates were selected with suitable multiplicity trigger conditions requiring at least two hits in TAPS or at least one hit in TAPS and two hits in the CB, derived from a fast cluster recognition encoder. A more detailed description of the detector setup and the running conditions can be found in \cite{thierry,nanova}. 

The $\eta^\prime$-mesons were identified via the $\eta^\prime \rightarrow \pi^{0} \pi^{0} \eta \rightarrow 6 \gamma$ decay channel, which has a branching ratio of 8.1\%. For the reconstruction of the $\eta^\prime$-meson, only events with at least 6 or 7 neutral hits were selected. Because of the competing channel   $\eta \rightarrow \pi^{0} \pi^{0} \pi^{0} \rightarrow  6 \gamma$ with the same final state, this reaction was also reconstructed and the corresponding events were rejected from the further analysis. In addition, only events with one combination of the 6 photons to two photon pairs with mass 110 MeV/c$^2 \le m_{\gamma\gamma} \le$ 160 MeV/c$^2$ close to the $\pi^{0}$ mass and one pair with  mass 500 MeV/c$^2\le m_{\gamma\gamma} \le $600 MeV/c$^2$ close to the $\eta$ mass were analyzed further. 

\begin{figure}[h!]
 \resizebox{0.8\textwidth}{!}{
    \includegraphics[height=0.4\textheight]{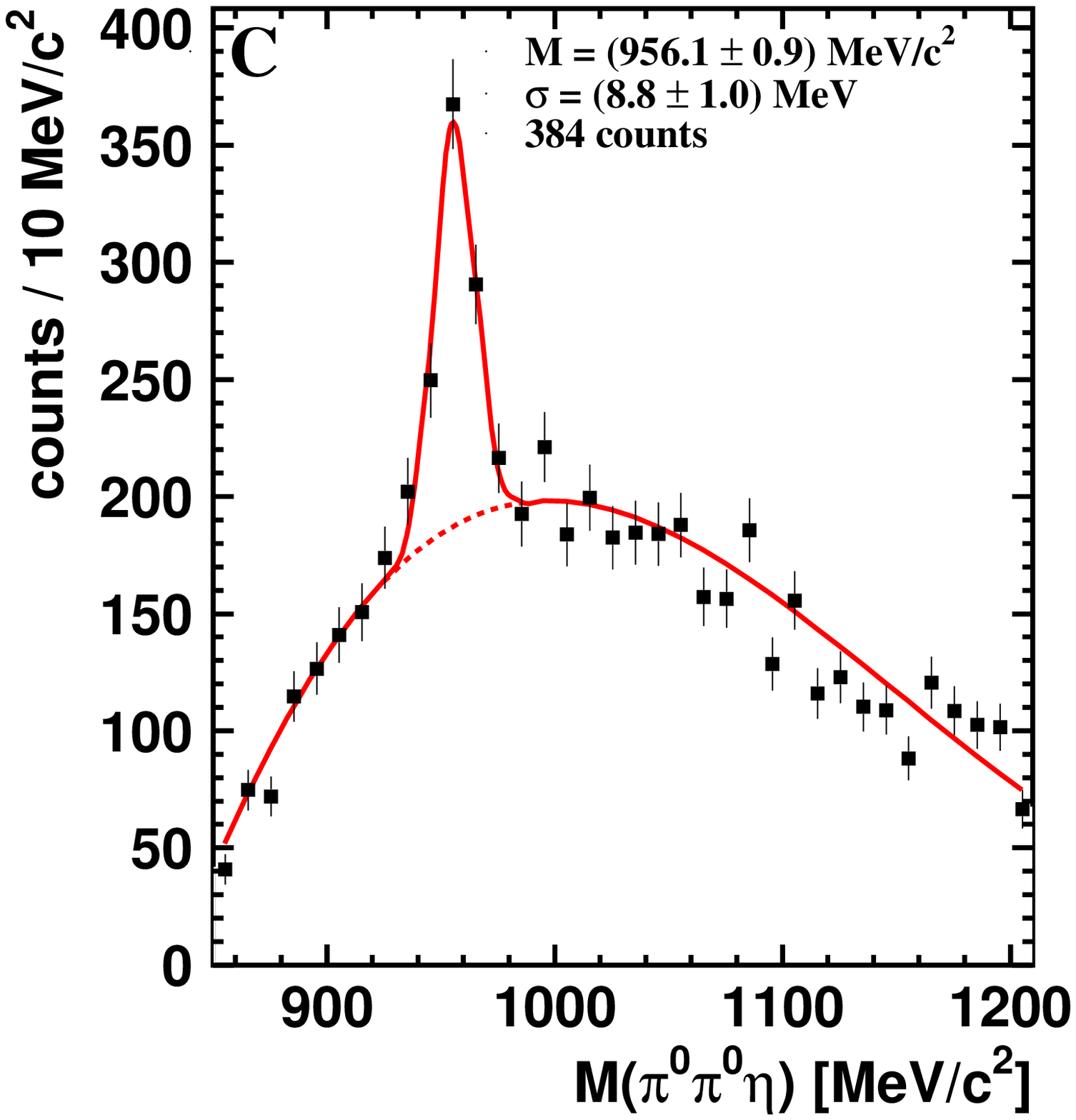}\includegraphics[height=0.4\textheight]{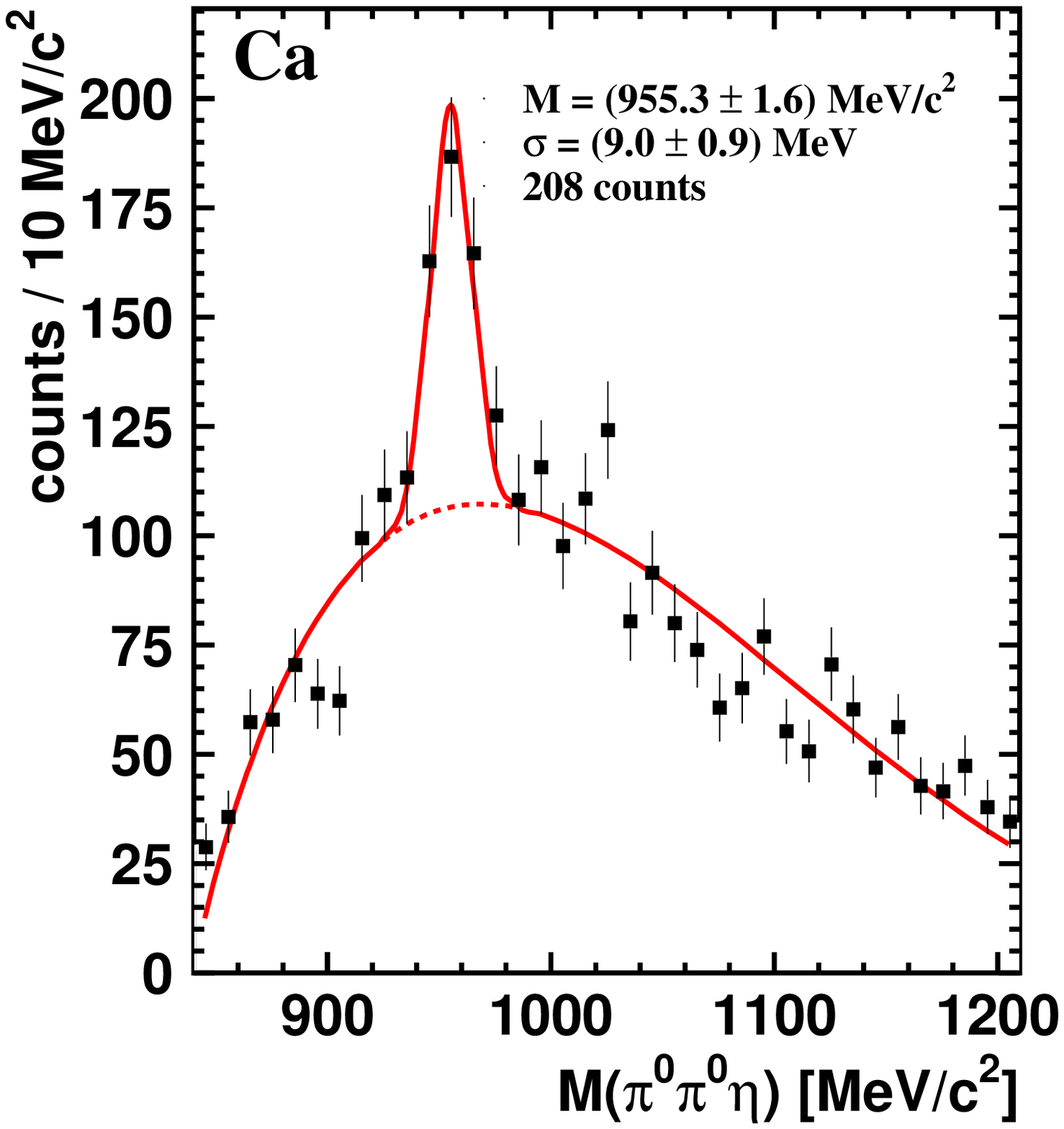}}
 \resizebox{0.8\textwidth}{!}{%
    \includegraphics[height=0.4\textheight]{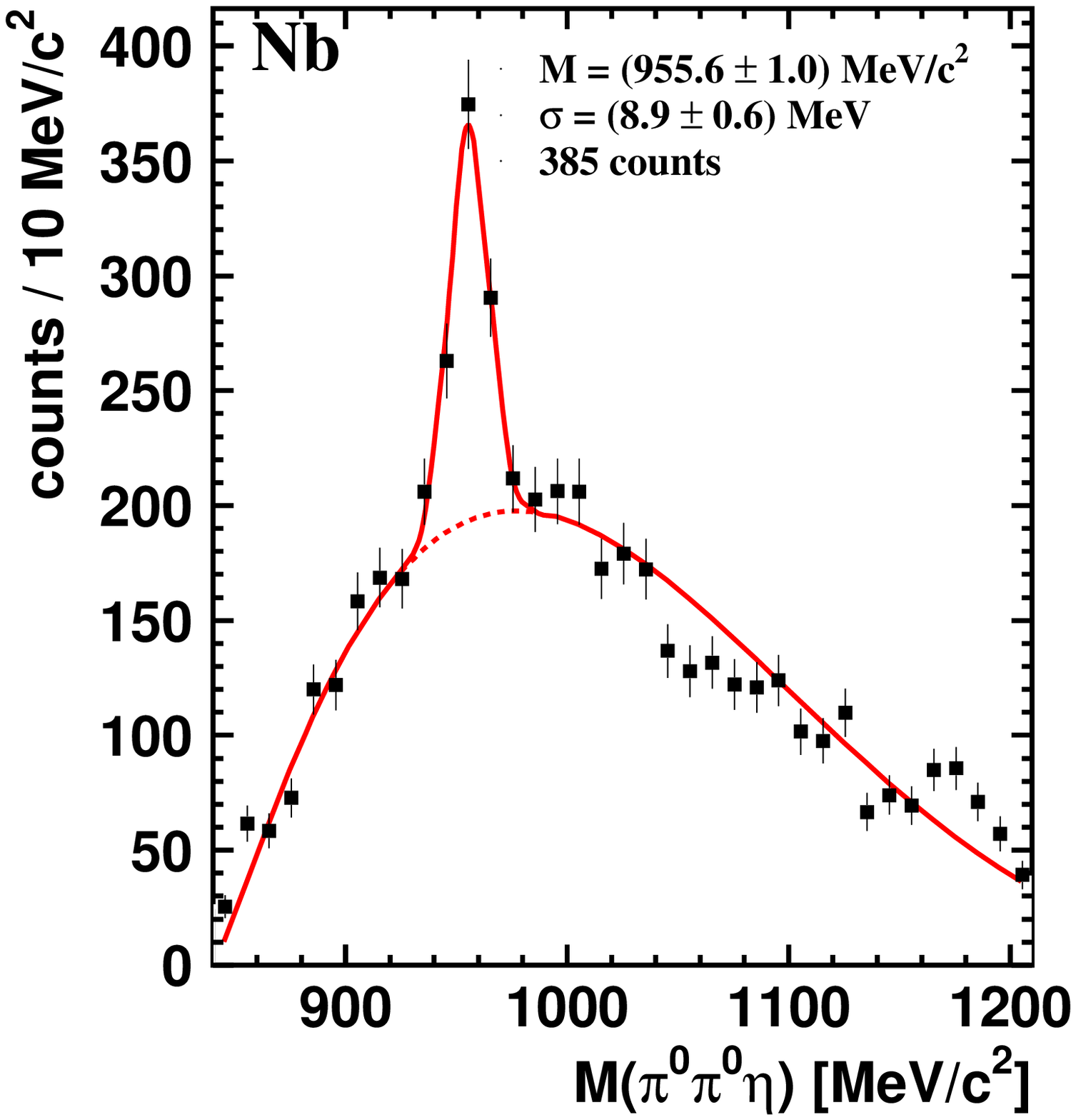}\includegraphics[height=0.4\textheight]{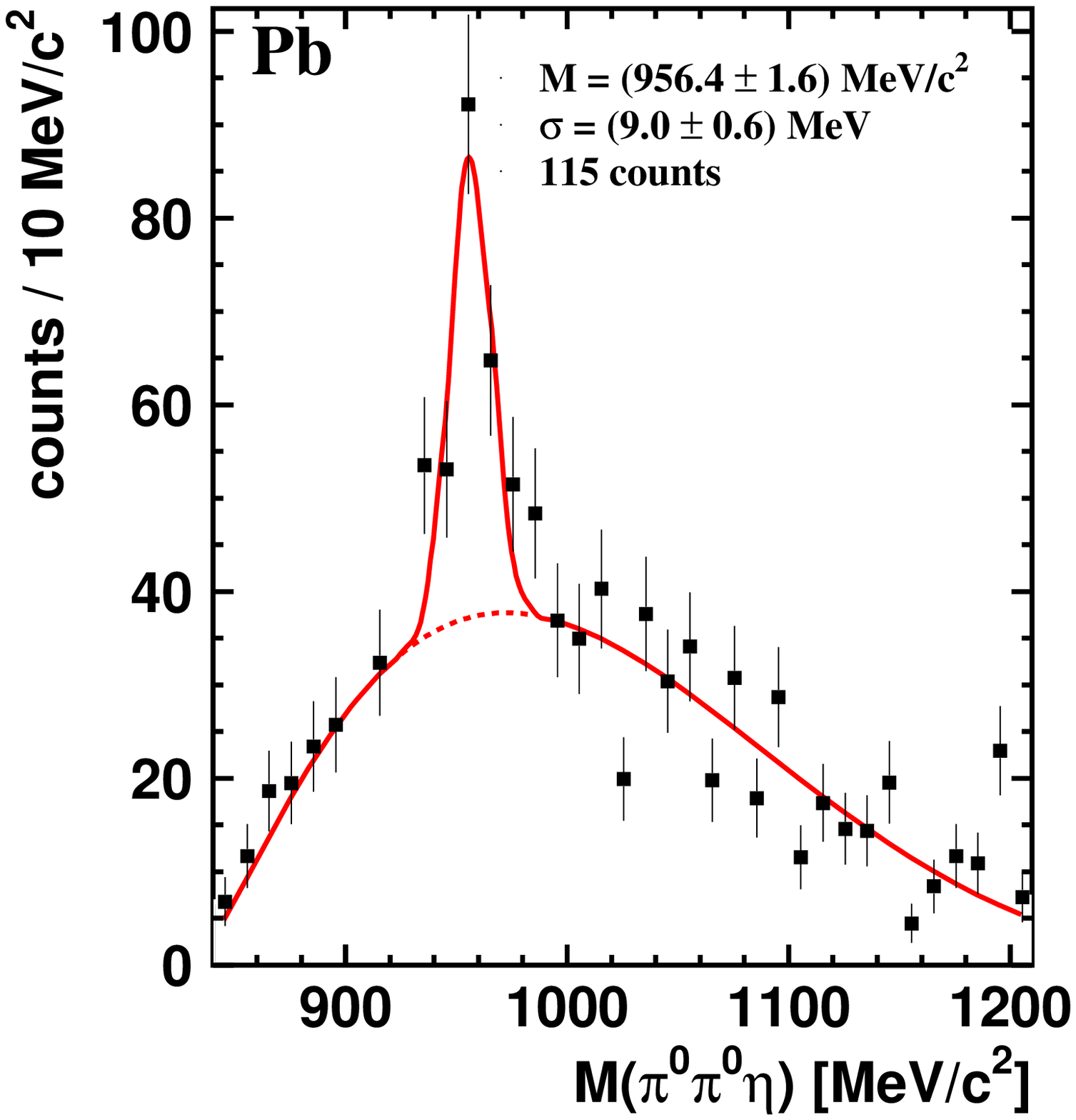}}
\caption{Invariant mass spectrum of $\pi^{0}\pi^{0}\eta$ for ${}^{12}\textrm{C}, {}^{40}\textrm{Ca}, {}^{93}\textrm{Nb}$ and ${}^{208}\textrm{Pb}$ targets for the incident photon energy range 1500 - 2200 MeV. The solid curve is a fit to the spectrum. Only statistical errors are given. See text for more details.}
\label{fig:invmass}
\end{figure}

The $\pi^0\pi^0\eta$ invariant mass spectra for different targets and the incident photon energy range from 1500 -  2200~MeV are shown in Fig.~\ref{fig:invmass}. 
The spectra were fitted with a Gaussian and a background function $f(m) = a \cdot (m-m_1)^{b} \cdot (m-m_2)^{c} $.  Alternatively, the signals  were fitted by a function allowing for low mass tails as in \cite{nanova} and the background shape was also fitted with a polynomial. Variations in the determined 
$\eta^\prime$ yields were of the order of 10-15 \% and represent the systematic errors of the fitting procedures. For the cross section determination the acceptance for the detection of an $\eta^\prime$-meson in the inclusive $\gamma $A$ \rightarrow \eta^\prime$ + X reaction was simulated as a function of its kinetic energy and emission angle in the laboratory frame, as described in \cite{thierry}. Thereby a reaction-model independent acceptance corrections is obtained which is applied event-by event to the data. Particles were tracked through the experimental setup using GEANT3 with a full implementation of the detector system, as described in more detail in \cite{Igal}. However, since only cross section ratios are presented, systematic errors in the acceptance determination tend to cancel. The photon flux through the target was determined by counting the photons reaching the $\gamma$ intensity detector at the end of the setup in coincidence with electrons registered in the tagger system. As discussed in \cite{Nanova_K*}, systematic errors introduced by the photon flux determination are estimated to be about 5-10 $\%$. Systematic errors of $\approx 10\%$ arise from uncertainties in the effective number of participating nucleons seen by the incident photons due to photon shadowing (see \cite{bianchi}). The different sources of systematic errors are summarized in Table 
\ref{tab:syst}. The total systematic errors in the determination of  the transparency ratios and of quantities derived from them are of the order of 20$\%$.
\begin{table}[h!]
\centering
\begin{footnotesize}
\begin{tabular}{|c|c|}
\hline
fits & $\approx 10-15\%$\\
acceptance & $\approx 5\%$\\
photon flux & 5-10 $\%$\\
photon shadowing & $\approx 10\%$\\
\hline
total& $\approx 20\%$\\
\hline
\hline
\end{tabular}
\end{footnotesize}
\caption{sources of systematic errors}
\label{tab:syst}
\end{table}

\subsection{Results and discussion} 
\begin{figure*} 
 \resizebox{1.\textwidth}{!}{
\includegraphics[height=0.45\textheight]{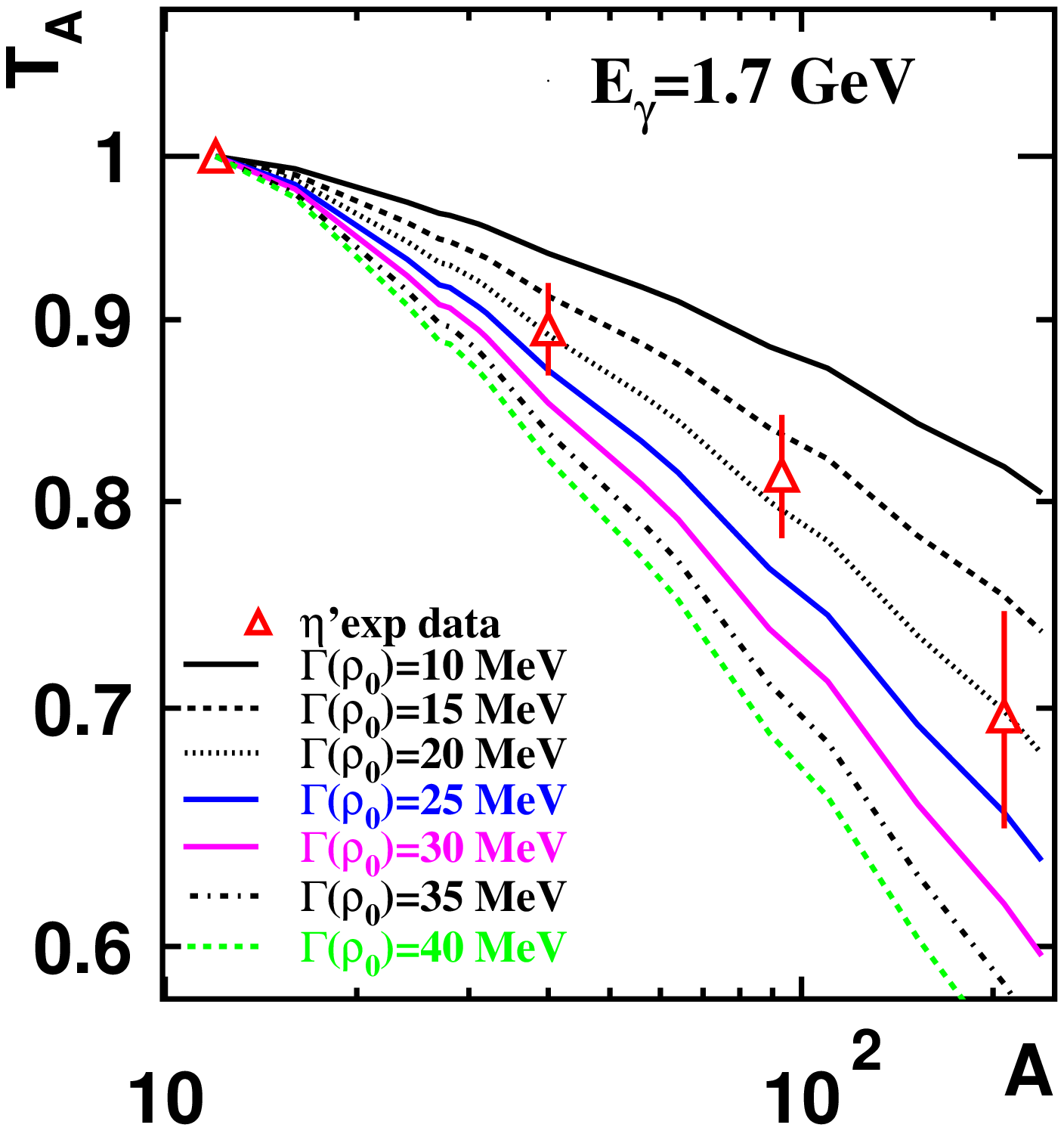}\includegraphics[height=0.45\textheight]{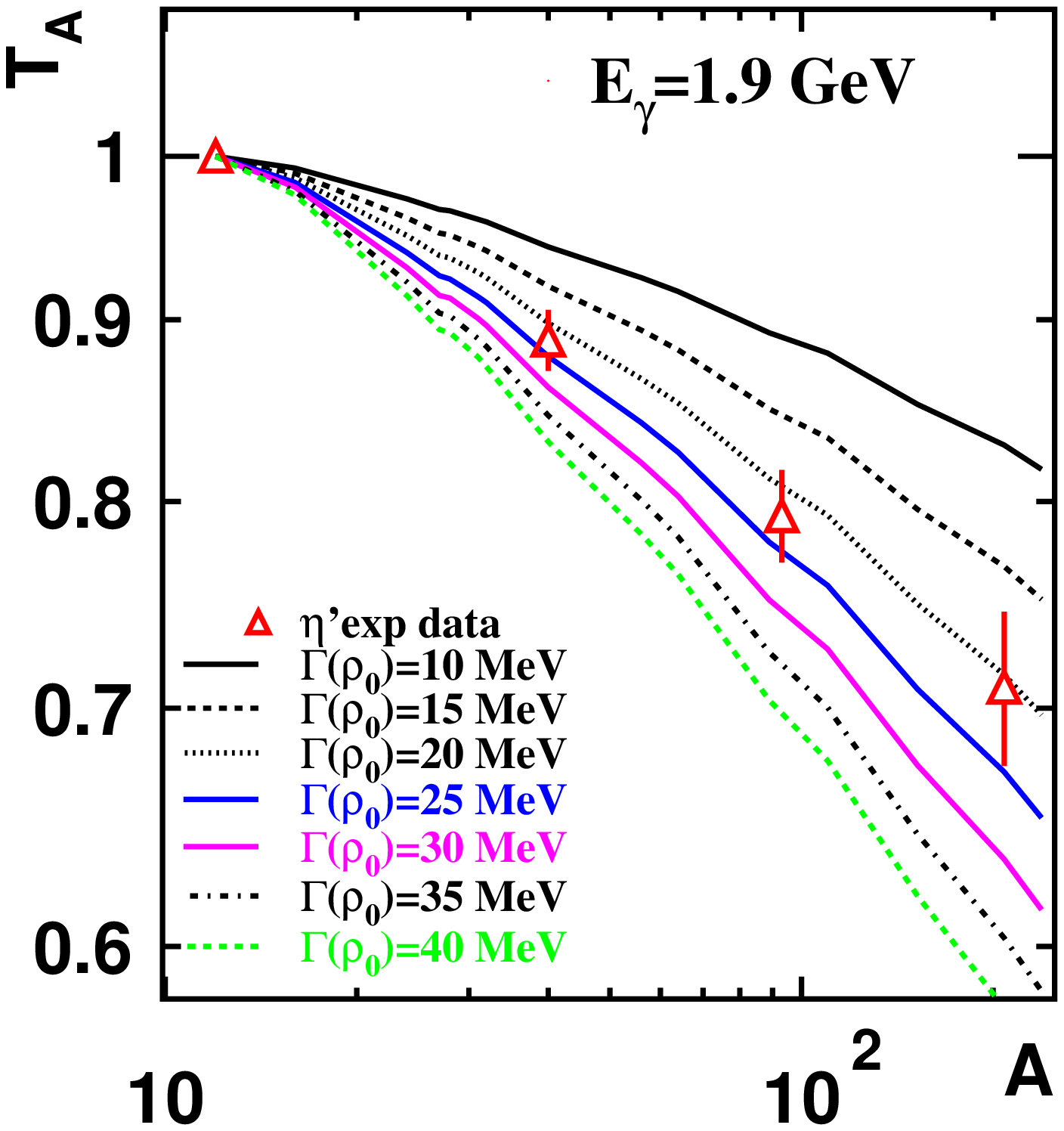}\includegraphics[height=0.45\textheight]{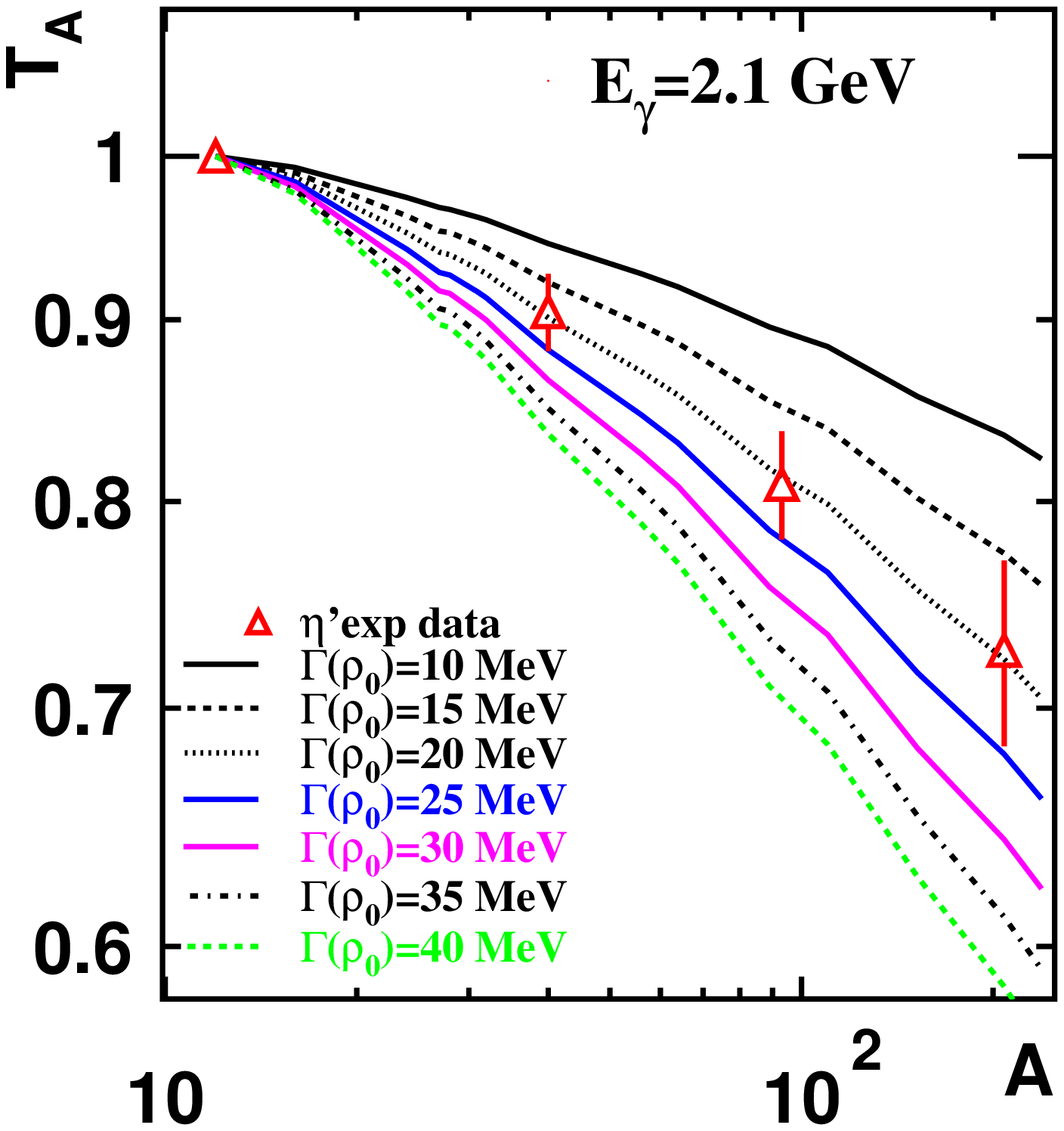}}
\caption{Transparency ratio relative to that of $^{12}$C, $T_A=\tilde{T}_A/\tilde{T}_{12}$, as a function of the nuclear mass number $A$, for different in-medium widths of the $\eta^\prime$ at three different incident photon energies. Only statistical errors are shown. The systematic errors are of the order of 20$\%$ but tend to partially cancel since cross section ratios are given.}
\label{fig:ratio}
\end{figure*}

Cross sections were measured for the four targets and the resulting transparency ratios were normalized to carbon, according to Eq.(\ref{eq:trans_ratio}). The transparency ratio as a function of the nuclear mass number $A$ is shown in Fig.(\ref{fig:ratio})
for three different incident photon energy bins, namely: 1600-1800~MeV, 1800-2000~MeV and 2000-2200~MeV. These curves are calculated using eqs. (1) to (7) for different values of the in-medium width $\Gamma_{\eta^\prime} (\rho_0)$ of the $\eta^\prime$-meson in eq. (7), ranging from 10~MeV to 40~MeV. The magnitude of $\Gamma$ at $\rho_0$, the normal nuclear matter density, is used in what follows when we refer to the in-medium width. 

Best agreement with the experimental data is obtained for an in-medium width of the $\eta^\prime$ -meson of 15-25~MeV.  Assuming the low density approximation
\begin{equation}
\Gamma = \rho_{0} \sigma_{\rm inel} \beta, \label{eq:Gamma-sigma}
\end{equation}
with
\begin{equation}
\beta = \frac{p_{\eta^\prime}}{E_{\eta^\prime}}
\end{equation}
in the laboratory
and taking the average $\eta^\prime$ recoil momentum of 1.05 GeV/c into account, an inelastic cross section of $\sigma_{\rm inel} \approx$ 6-10~mb is deduced.

This value is consistent with the result of a Glauber model analysis. Within this approximation an expression for the transparency ratio has been derived in \cite{MM} 
\begin{equation}
T_A=\frac{\pi R^{2}}{A \sigma_{\eta^\prime N}} \left\{1+\left(\frac{\lambda}{R}\right) \exp\left[-2\frac{R}{\lambda} \right ] +\frac{1}{2}\left(\frac{\lambda}{R}\right)^{2} \left(\exp \left [ -2\frac{R}{\lambda} \right ]-1\right) \right\} \label{MM_fit}
\end{equation}
where $\lambda= (\rho_{0} \sigma_{\eta^\prime N})^{-1} $ is the mean free path of the $\eta^\prime$-meson in a nucleus with density $\rho_{0}$=0.17 fm$^{-3}$ and radius $R=r_{0} A^{1/3}$ with $r_{0}=1.143$ fm.  Fitting  
\begin{figure}[h!]
 \resizebox{0.7\textwidth}{!}{
    \includegraphics[height=0.5\textheight]{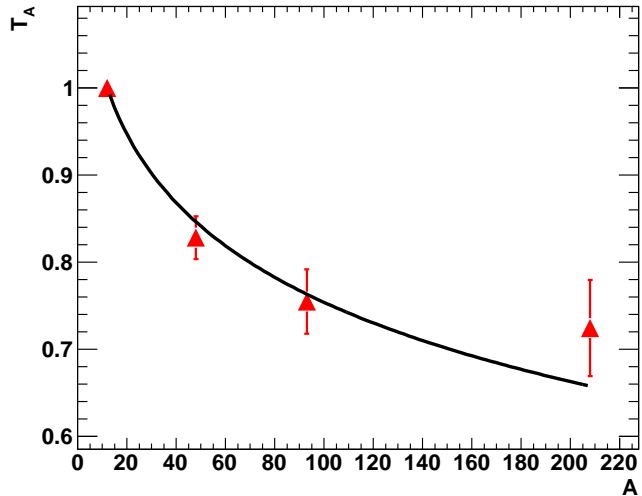}}
\caption{The transparency ratio for $\eta^\prime$-mesons as a function of the nuclear mass number $A$ for the full incident photon energy range of 1500 - 2200 MeV. The solid curve is a fit to the data using expression in Eq.(\ref{MM_fit}).}
\label{fig:MM_fit}
\end{figure}
this expression to the $\eta^\prime$ transparency ratio data shown in Fig.~\ref{fig:MM_fit} an in-medium $\eta^\prime$N inelastic cross section of $\sigma_{\eta^\prime N}$ = (10.3$\pm$1.4) mb is obtained. 

So far, in order to determine $\sigma_{\rm inel}$ we have assumed that the $\eta^\prime$ absorption process is dominated by one-body absorption. In \cite{Nagahiro1} two-body absorption mechanisms have also been evaluated; close to threshold results have been obtained in terms of the unknown $\eta^\prime$~N scattering length. Although the energies of the $\eta^\prime$ are on average higher in the present experiment, the results of \cite{Nagahiro1} are used to estimate the uncertainties:  if $\mid a_{\eta^\prime N} \mid$ is of the order of
0.1 fm, the $\eta^\prime$ width at $\rho_0$ is of the order of 2 MeV,
and only 6$\%$ of it is due to two body absorption mechanisms. Obtaining
a width as large 20 MeV, as found here, would require
values of  $\mid a_{\eta^\prime N} \mid$ of the order of 0.75 fm, in
which case the contributions of the one-body and two-body absorption
mechanisms turn out to be similar. We consider this to be a rather
extreme situation, providing a boundary for the determination of
$\sigma_{\rm inel}$. In this case the density dependence of the width
would be given by $\Gamma^{1+2}_{\eta^\prime}
(\rho)=\Gamma^{1+2}_{\eta^\prime}(\rho_0) [\rho/\rho_0+
(\rho/\rho_0)^2]/2. $ An explicit calculation using this density dependence gives rise to very similar
curves for different values of $\Gamma^{1+2}_{\eta^\prime}(\rho_0)$ as in Fig.~\ref{fig:ratio}, only displaced slightly upwards. 
The best agreement
with the data is then found for $\Gamma^{1+2}_{\eta^\prime}(\rho_0)=$
17-27 MeV. The similarity of this value to the
width of 15-25 MeV obtained from the one-body absorption analysis
indicates that $\eta^\prime$ absorption occurs
in regions of full density ($\rho \simeq  \rho_0$) where both one- and
two-body mechanisms contribute
equally with the density functional chosen. The presence of two-body
$\eta^\prime$ absorption processes makes the $\eta^\prime$ inelastic cross
section smaller than the value extracted assuming all $\eta^\prime$
absorption to be of one-body
type. In the extreme case analyzed here, only half of the width comes
from one-body absorption mechanisms and, consequently, the inelastic
cross section gets reduced by a factor of two to the value
$\sigma_{\rm inel}=$ 3-5 mb. In summary, while for the width at
$\rho_0$ we determine a range of about 15-25~MeV, the range of $\sigma_{\rm inel}$ values gets enhanced to about 3 - 10~mb to account for
uncertainties arising from the unknown strength of two-body absorption processes.

\begin{figure}[h!]
 \resizebox{0.5\textwidth}{!}{
    \includegraphics[height=0.5\textheight]{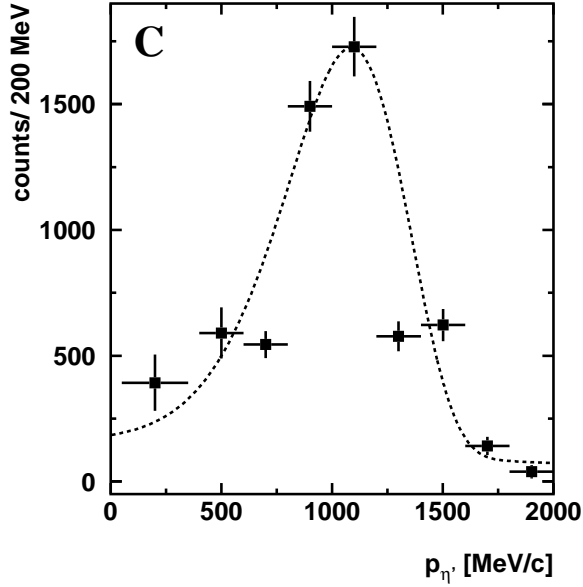}}
\caption{Momentum distribution of $\eta^\prime$-mesons produced off a C target for $E_{\gamma}^{beam}$ = 1500 - 2200 MeV. The dashed curve is to guide the eye.}
 \label{fig:mom}
\end{figure}

\begin{figure}[h!]
 \resizebox{1.\textwidth}{!}{
    \includegraphics[height=0.8\textheight]{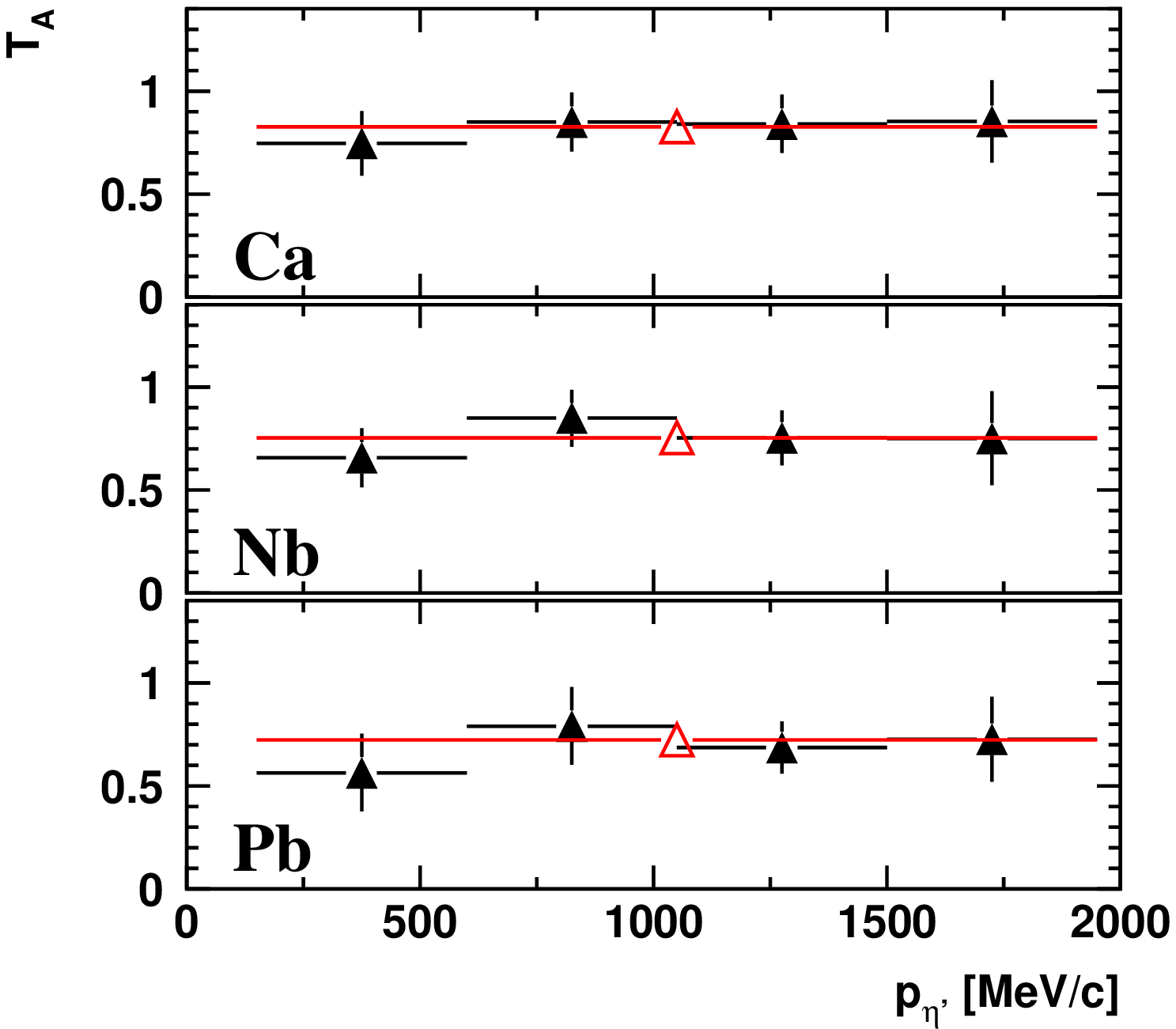}\includegraphics[height=0.65\textheight]{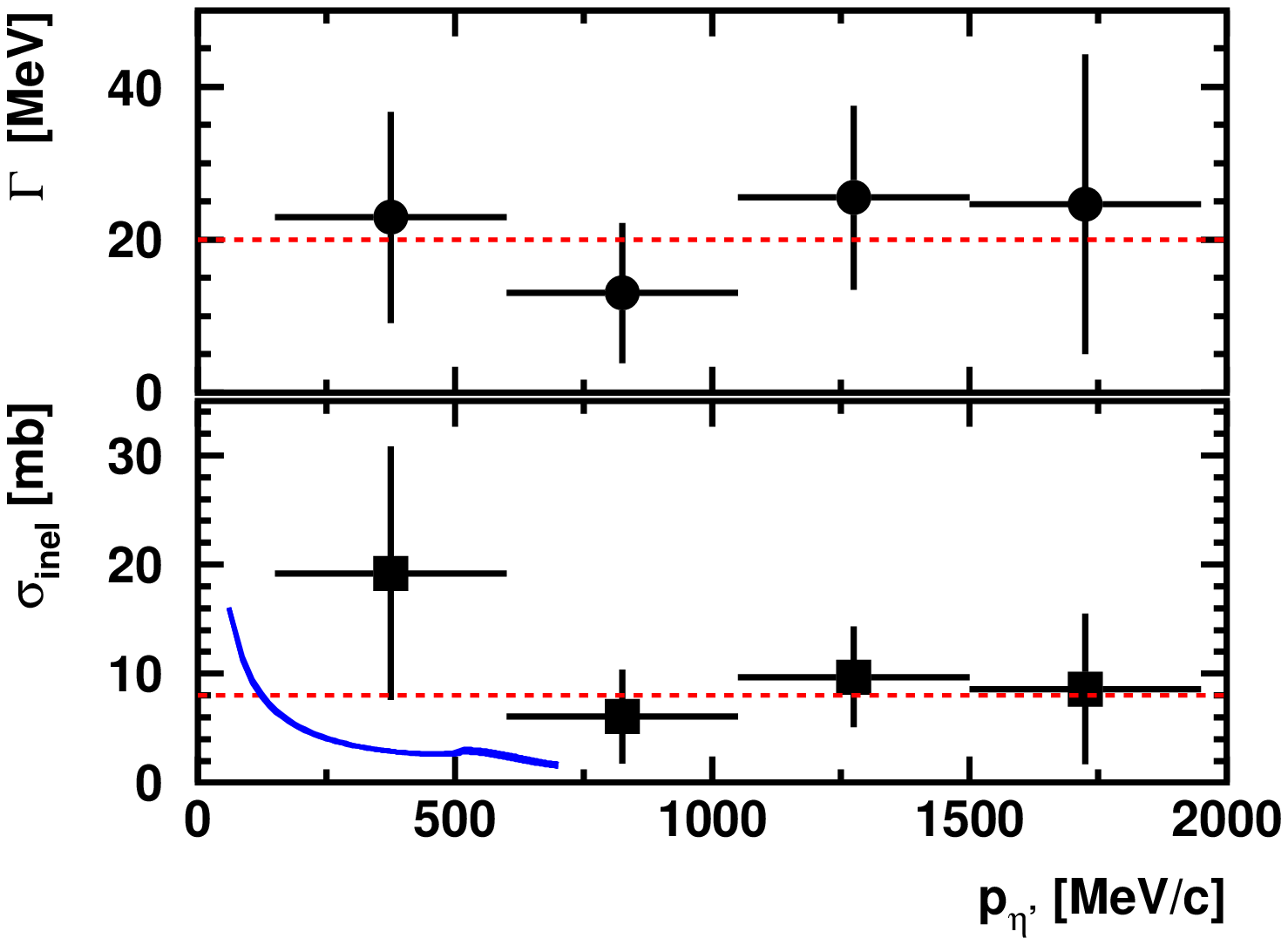}}
\caption{(Left) Transparency ratio for the $\eta^\prime$-meson normalized to C for three different targets: Ca, Nb, and Pb and four bins in $\eta^\prime$ momentum (full triangles) for the full incident photon energy range from 1500 - 2200~MeV. The open triangles show the values when integrated over all momenta and energies. (Right) The in-medium width (upper panel) and inelastic cross section (lower panel) as a function of the $\eta^\prime$ momentum.
For comparison, the theoretical predictions for $\sigma_{\rm inel}$ \cite{Oset:2010ub} are shown by a blue (solid) curve. }
 \label{fig:ta_p}
\end{figure}

The momentum distribution of the $\eta^\prime$-mesons produced off a C-target is shown in Fig.~\ref{fig:mom}. The distribution peaks at about 1.1 GeV/c 
which is close to the average momentum of 1.05~GeV/c. 
The transparency ratio has also been determined for four bins in  $\eta^\prime$ momentum to study the momentum dependence. After correcting for the momentum and target dependent $\eta^\prime$ acceptance, Fig.~\ref{fig:ta_p} (Left) exhibits only a weak variation of the transparency ratio with the $\eta^\prime$ momentum. 
 Having determined the transparency ratio for different momentum bins also the in-medium inelastic cross section and the in-medium width can be derived as a function of the $\eta^\prime$ momentum, applying fits as in Fig.~\ref{fig:MM_fit} for each momentum bin separately. Within the errors no strong variation with momentum is observed as shown in Fig.~\ref{fig:ta_p} (Right).

This  indicates that two-step processes do not seem to play an important role in the photoproduction of $\eta^\prime$-mesons in the photon energy regime studied. This is an important observation because Eq.(\ref{eq:Gamma-sigma}) can only be applied to extract an inelastic cross section if two-step processes can be neglected. Otherwise the measured transparency ratio would reflect a convolution of secondary production and absorption in nuclei. In two-step processes where e.g. a pion is produced in the initial step by the incoming photon and the $\eta^\prime$-meson is then subsequently produced in a pion-induced reaction on another nucleon there is less energy available for the final state meson. This would shift the 
$\eta^\prime$ yield towards lower energies and lead to an enhancement of the transparency ratio at low $\eta^\prime$ momenta.

Because of the near constancy of $\Gamma$ one would expect (see Eq. (9)) a rise of $\sigma_{\rm inel}$ towards lower $\eta^\prime$ momenta, as indicated by the data in the lower panel of Fig.5 (right). An increase of $\sigma_{\rm inel}$ for low $\eta^\prime$ momenta has in fact been predicted in \cite{Oset:2010ub}, rather independent of the $\eta^\prime$ scattering length. The theoretical predictions follow qualitatively the trend of the data and may even be compatible with the experimental results, allowing for the large systematic uncertainties in the determination of $\sigma_{\rm inel}$ due to the unknown strength of two-body absorption processes, discussed above.

\begin{figure}[h!]
 \resizebox{1.\textwidth}{!}{
    \includegraphics[height=0.5\textheight]{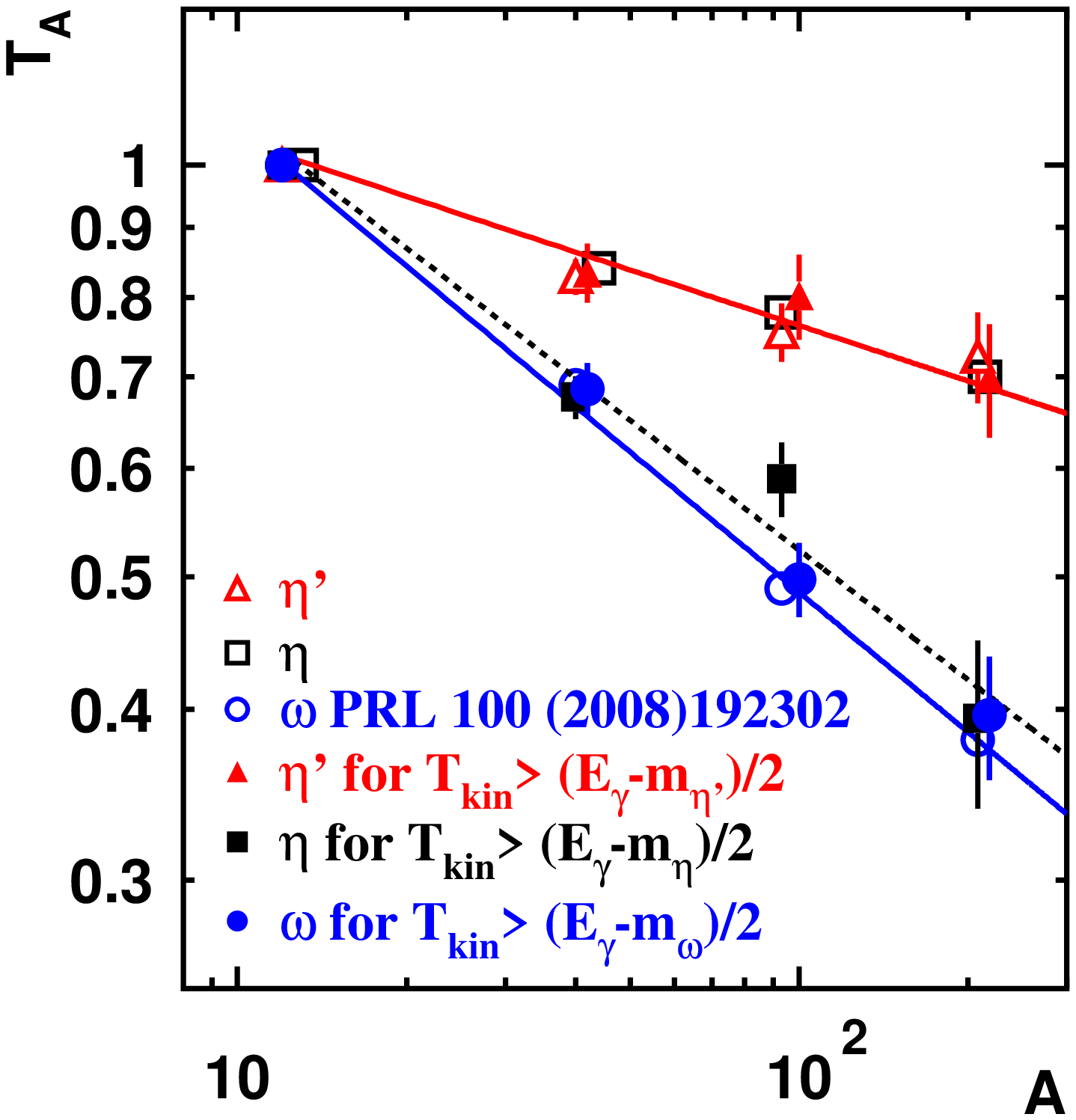}\includegraphics[height=0.5\textheight]{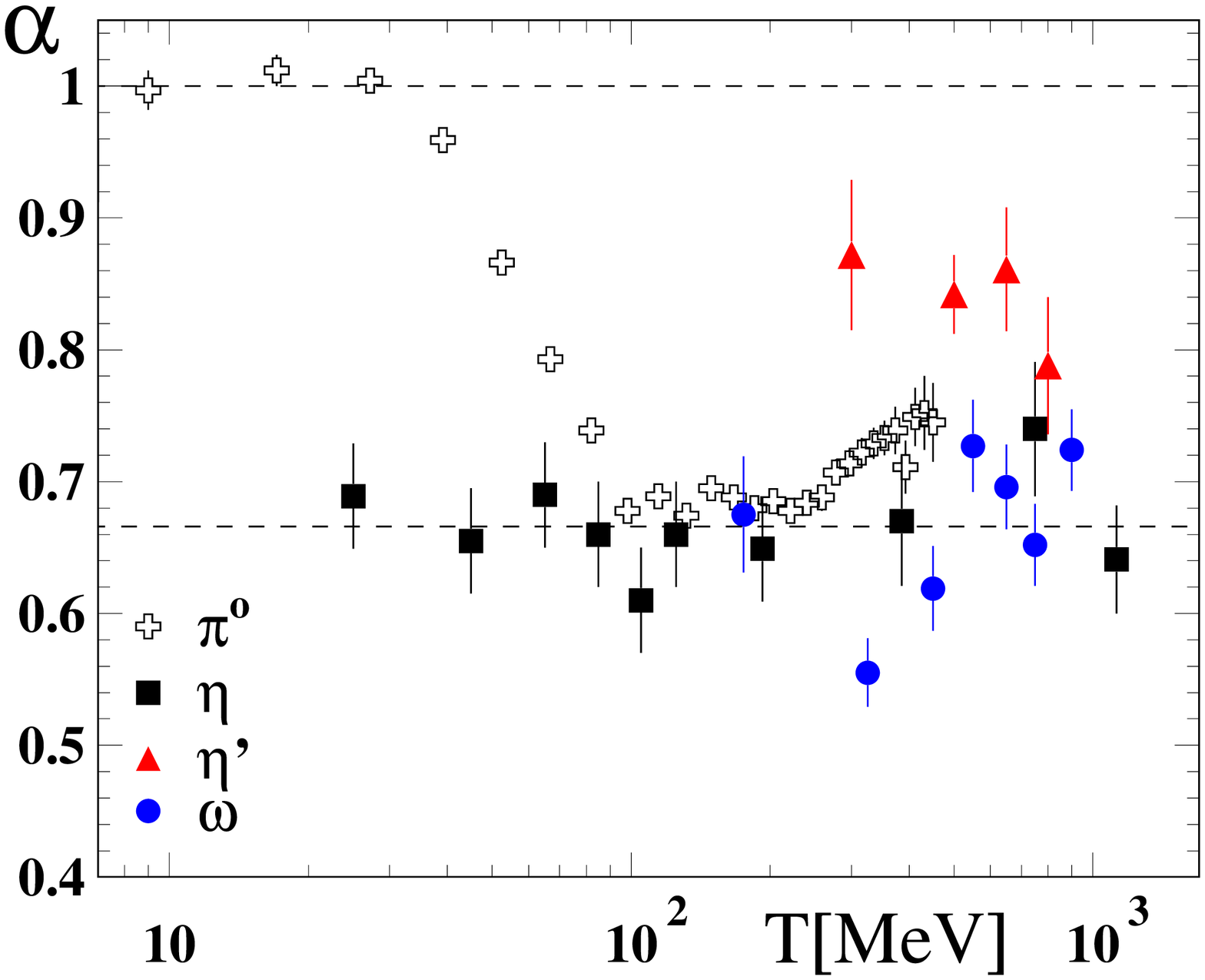}}
\caption{(Left) Transparency ratio for different mesons - $\eta$(squares), $\eta^\prime$(triangels) and $\omega$(circles) as a function of the nuclear mass number $A$.  The transparency ratio with a cut on the kinetic energy for the respective mesons is shown with full symbols. The incident photon energy is in the range 1500 to 2200~MeV. The solid lines are fits to the data. Only statical errors are shown. The impact of photon shadowing on the determination of the transparency ratio has been taken into account for the $\eta^\prime$ meson, but has not been corrected for in the published data for the other mesons. (Right) $\alpha$ parameter dependence on the kinetic energy T of the meson compared for $\pi^0$ ~\cite{bernd}, $\eta$ ~\cite{robig, thierry}, $\eta^\prime$ and $\omega$ (\cite{Kotulla}, this work). This figure is an updated version of a figure taken from \cite{thierry}. } 
 \label{fig:ta}
\end{figure}

In Fig.~\ref{fig:ta} the results for the $\eta^\prime$-meson are compared to transparency ratio 
measurements for the $\eta$ \cite{thierry} and $\omega$ meson \cite{Kotulla}. In this comparison it should be noted that - in contrast to the present work - the impact of photon shadowing on the transparency ratio had not been taken into account in earlier publications. The data are shown for the full kinetic energy range  of recoiling mesons (open symbols) as well as for the fraction of high energy mesons (full symbols) selected by the constraint
\begin{equation}
T_{kin} \ge (E_{\gamma} -m)/2. \label{cut}
\end{equation}
 Here, $E_{\gamma} $ is the incoming photon energy and $T_{kin} $ and $m$ are the kinetic energy and the mass of the meson, respectively. As discussed in \cite{thierry}, this cut suppresses meson production in secondary reactions. Fig.~\ref{fig:ta} (left) shows that within errors this cut does not change the experimentally observed transparency ratios for the $\omega$-meson and $\eta^\prime$-meson while there is a significant difference for the $\eta$ meson. For the latter, secondary production processes appear to be more likely in the relevant photon energy range because of the larger available phase space due to its lower mass (547~MeV/c$^2$) compared to the $\omega$ (782~MeV/c$^2$) and $\eta^\prime$ (958~MeV/c$^2$) meson. The spectral distribution of secondary pions, which falls off to higher energies, together with the cross sections for pion-induced reactions favor secondary production processes in case of the $\eta$-meson: 3 mb at $p_{\pi} \approx$ 750 MeV/c in comparison to 2.5 mb at $p_{\pi} \approx $ 1.3 GeV/c for the $\omega$-meson and 0.1 mb at $p_{\pi} \approx$ 1.5 GeV/c for the $\eta^\prime$-meson, respectively \cite{Landolt-Boernstein}. In addition, $\eta$-mesons may be slowed down through rescattering with secondary nucleons, which can be enhanced by the S$_{11}$(1535) excitation. According to Fig.~\ref{fig:ta} (left) the $\eta^\prime$-meson shows a much weaker attenuation in normal nuclear matter than the $\omega$ and $\eta$-meson, which exhibit a similarly strong absorption after suppressing secondary production effects in case of the $\eta$-meson.
  
An equivalent representation of the data can be given by parameterizing the observed meson production cross sections by $\sigma(A) = \sigma_0 A^{\alpha(T)}$ where $\sigma_0$ is the photoproduction cross section on the free nucleon and $\alpha$ is a parameter depending on the meson and its kinetic energy. The value of $\alpha \approx$ 1 implies no absorption while $\alpha \approx $ 2/3 indicates meson emission only from the nuclear surface and thus implies strong absorption. All results are summarized in Fig.~\ref{fig:ta} (right) and additionally compared to data for pions \cite{bernd}. For low-energy pions, $\alpha \approx 1.0$ because of a compensation of the repulsive s-wave interaction by the attractive p-wave  $\pi N$ interaction. This value drops to $\approx$ 2/3 for the $\Delta$ excitation range and slightly increases for higher kinetic energies. After suppressing secondary production processes by the cut (Eq.(\ref{cut})) the $\alpha$ parameter for the $\eta$-meson is close to 2/3 for all kinetic energies, indicating strong absorption \cite{thierry}. For the $\omega$-meson the $\alpha$ values are also close to 2/3. The weaker interaction of the $\eta^\prime$-meson with nuclear matter is quantified by $\alpha = 0.84\pm 0.03$ averaged over all kinetic energies.

\section{Conclusions}
The transparency ratios for $\eta^\prime$-mesons measured for several nuclei
deviate sufficiently from unity to allow an extraction of the  $\eta^\prime $ width in the nuclear medium, and an approximate inelastic cross section for $\eta^\prime N$ at energies around $\sqrt s \approx$ 2.0~GeV. We find $\Gamma \approx  15-25$~MeV $\cdot \rho/\rho_0$ roughly, corresponding to an inelastic $\eta^\prime N$ cross section of $\sigma_{\rm inel} \approx$ 6-10~mb. If inelastic and two-body absorption processes were equally strong the inelastic cross section would be reduced to  $\sigma_{\rm inel} \approx$ 3-5~mb. Despite of the uncertainties and approximations involved in the determination of $\sigma_{\rm inel}$, this is the first experimental measurement of this cross section. A comparison to photoproduction cross sections and transparency ratios measured for other mesons ($\pi,\eta,\omega$) demonstrates the relatively weak interaction of the $\eta^\prime$-meson with nuclear matter. Regarding the observability of $\eta^\prime$ mesic states the measured in-medium width of $\Gamma \approx  15-25$~MeV  at normal nuclear matter density would require a depth of about 50 MeV or more for the real part of the $\eta^\prime$ - nucleus optical potential.

\section*{Acknowledgments}
We thank the scientific and technical staff at ELSA for their important contribution to the success of the
experiment. We acknowledge detailed discussions with U. Mosel. 
This work was supported financially by the {\it Deutsche Forschungsgemeinschaft}  through SFB/TR16 and
partly supported by DGICYT contracts  FIS2006-03438,
FPA2007-62777, the Generalitat Valenciana in the program Prometeo and 
the EU Integrated Infrastructure Initiative Hadron Physics
Project  under Grant Agreement n.227431. The Basel group
acknowledges support from the {\it Schweizerischer Nationalfonds}.

\end{document}